\documentstyle[aps,eqsecnum,preprint,prbbib]{revtex}
\begin{document}
\draft
\title{Numerically Implemented Perturbation Method for
the Nonlinear Magnetic Moment of an Anisotropic Superconductor} 
\author{Igor \v{Z}uti\'c and Oriol T. Valls}
\address{School of Physics and Astronomy and Minnesota Supercomputer Institute
\\ University of Minnesota \\
Minneapolis, Minnesota 55455-0149}
\maketitle 
\vspace{90 mm}
7808, 78A25, 65N06 \\
Maxwell-London electrodynamics, superconductivity, magnetic moment, 
perturbation method, nonlinear partial differential equations,          
finite-differences, spheroidal geometry. 
\pagebreak
\\
Nonlinear Magnetic Moment of a 3D Superconductor \\
\\
\\
Oriol T. Valls \\
School of Physics and Astronomy and Minnesota Supercomputer Institute
\\ University of Minnesota \\
116 Church Street S.E. \\
Minneapolis, Minnesota 55455-0149 \\
email: otvalls@maroon.tc.umn.edu \\
fax: (612) 624-4578
\pagebreak
\begin{abstract}
We present a method to compute the magnetic moment of a bulk,          
finite-size, three-dimensional, anisotropic superconductor. Our numerically  
implemented perturbative procedure is based on a solution of the  nonlinear  
Maxwell-London electrodynamic equations, where we include the nonlinear 
relation between current and gauge invariant velocity. The method exploits  
the small ratio of  
the finite penetration depths to the sample size. We show how to treat 
the open boundary conditions over an infinite domain and the continuity 
requirement at the interface. We demonstrate how our method 
substantially reduces the computational work required, and discuss its 
implementation to an oblate spheroid. The numerical 
solution is obtained from a finite-difference method. 
We briefly discuss the relevance of this work to similar problems in 
other fields.  
\end{abstract}
\pagebreak
\section{Introduction}
\label{intro}
A large number of problems in electrodynamics and related areas such as
fluid dynamics, involve
the solution of partial differential equations for certain fields
inside and outside a finite region of a definite geometrical shape. For
many common geometries, and when the boundary conditions are simple (e.g.
fields or their normal derivatives vanishing at the boundaries) the
solution can be found, often with ease, from analytical or simple
numerical methods. However, for more complicated situations where one
has less trivial boundary conditions, 
or when the equations are made more complicated by the presence of
nonlinearities,
analytical methods may be unavailable and numerical techniques encounter
serious difficulties.

One of these situations pertains to the electrodynamics of a superconducting
sample of finite size. It is well-known that in the limit
where the electromagnetic fields do not penetrate the sample the problem
can be rather easily solved [\onlinecite{landau}].
However, this is hardly ever the case of
interest: the physical information one obtains in experiments comes in fact
from the penetration of the fields inside the sample, characterized 
by penetration depths which, although small, cannot be neglected.

Consider a superconductor that occupies a bounded, macroscopic region 
$\Omega \subset {\bf R}^3$,  in the  presence of an applied uniform
magnetic field, ${\bf H_a}$. 
For $H_a$ below some critical value, superconductors
are in the  so called Meissner regime [\onlinecite{kaper}], where the   
magnetic flux  
is expelled from the bulk of the sample. Their behavior is similar to that
of material which is both an ideal conductor and an ideal diamagnet.
The applied magnetic field generates a resistance-free current which
produces a magnetic field that opposes $H_a$. As a consequence,
everywhere except very close to the interface (within a few penetration 
depths), the magnetic field vanishes:
 this is known as the Meissner effect.
Except for the most
trivial geometries such as infinite slabs or isotropic
spheres, the relevant boundary value problem becomes then numerically
very awkward: basically one is faced with solving the appropriate
electrodynamic equations in the entire space,
not just in $\Omega$,
while the most important  variation of the fields takes place in a very
thin boundary layer in $\Omega$.

This question has come recently to the fore in the context of the study
of high temperature superconductors (HTSC's). Identifying the symmetry of
the paired electrons, the so called pairing state,
in these materials, would provide an important clue to the
still unknown mechanism responsible for superconductivity in
HTSC's. It turns out [\onlinecite{ysprl,sv,ys,zv}]
that careful measurements of certain electrodynamic properties in
superconductors, can
provide fingerprints for the nodal structure
of the order (gap) parameter [\onlinecite{ys}] (i.e. the points or lines in
the Fermi surface where it vanishes).  
The unconventional pairing states which are widely believed to exist in HTSC's,
produce nonlinearities in the electromagnetic response. As we shall 
see, the resistance-free current, in addition to the usual terms linear 
in the superfluid velocity, includes a small contribution for which the 
current-velocity relation is nonlinear.               
 These nonlinearities
then give rise [\onlinecite{sv,zv}], in the Meissner regime,
to a magnetic moment which has a small but detectable  {\it transverse}
 component, ${\bf m_\bot}$, 
perpendicular to the field ${\bf H_a}$ even when this is applied
along a direction of symmetry of the sample. This occurs when
the applied field lies in the $a-b$
crystallographic plane [\onlinecite{plakida}] (the $z$ axis is taken to be
along the $c$ crystallographic direction).
The angular dependence of the transverse component
$m_\bot$, as the crystal is rotated
about the $z$ axis, 
reflects directly the symmetry of the pairing state.
It is this quantity that has been experimentally
studied [\onlinecite{buan}] for purposes of the identification of the pairing  
state.

The physics of the situation
has been extensively discussed in  [\onlinecite{zv}],
and we deal here with  the mathematical and numerical
implications. The computation of the magnetic moment requires the
 solution of a problem
 of precisely the kind described in the previous
paragraphs. One must solve the appropriate electrodynamics, the
Maxwell-London equations  described in Section \ref{maxw}, for
all space, since at infinity the boundary condition requires that
 ${\bf H} \rightarrow {\bf H_a}$.
On $\Omega$ these equations contain, as we shall see, important and
nontrivial nonlinearities. A solution in the limit of zero penetration of
the fields in the sample is possible, but
it would be completely inadequate, 
since it would reflect only the geometry  and not
the detailed electromagnetic response of the superconductor.
On the other hand, the numerical solution for the nonlinear electrodynamics
in all space would be computationally demanding.

In this paper we present a discussion of the methods that we have developed
to obtain results [\onlinecite{zv}] for the nonlinear magnetic moment, including
both the longitudinal and transverse components, in HTSC's.             
 These methods involve the numerical implementation of a   
 perturbation scheme in which the small expansion
parameter is the appropriately defined ratio of an effective
penetration depth to a characteristic dimension of the superconductor.  
  We will show that this numerical implementation 
reduces the problem  essentially to that of finding
the numerical  solution in $\Omega$. 
In the region ${\bf R}^3\setminus \Omega$ outside the sample, one turns out  
to need  only a solution for the scalar 
Laplace equation with trivial Neumann boundary conditions. 
For a sufficiently symmetric $\Omega$, the form of the 
solution  can be obtained analytically, 
while for  some other cases a numerical solution would suffice.

In Section \ref{maxw} we discuss the nonlinear Maxwell-
London equations for a superconductor and show how they give rise to 
the magnetic moment.
 The geometrical shapes we have considered for $\Omega$                 
  (dictated by experimental considerations)
are discussed in Section \ref{geo}
where we introduce the general solution in ${\bf R}^3\setminus \Omega$.  
In Section \ref{comp} we discuss the computation of the magnetic moment and  
present the main result of this paper, the perturbative
method and its numerical implementation. Numerical considerations for 
the equations in $\Omega$ are described in the Section \ref{inside}.
In Section \ref{tests}, the equations are solved  for the 
previously discussed geometry, using a modified Gauss-Seidel 
relaxation, with the nonlinear terms (which are nonanalytic) included
through Picard's method.
We also illustrate the general ideas of the perturbation method. 
Finally, in Section \ref{future} we give  conclusions and   
guidelines for possible improvements and generalizations.
While our discussion is in the context of superconducting electrodynamics,
it will not escape the reader's notice that our procedures can be adapted
to many other problems in which one is faced with a small ``skin depth''
or similar parameter involving the penetration of fields or of their
derivatives inside a region, and the problem can be
more easily solved in the limit where this parameter vanishes.

\section{Maxwell-London Electrodynamics}
\label{maxw}
\subsection{Maxwell-London Equations}
\label{equ}
We begin by introducing the steady state Maxwell-London equations  
[\onlinecite{london,orlando,degennes}] that provide the framework to compute  
the field distributions we need in order to evaluate the magnetic moment.

As stated in the Introduction
we consider a superconductor in an applied uniform magnetic field  
${\bf H_a}$ that occupies a bounded simply connected region $\Omega$ 
$\subset$ ${\bf R}^3$ and at its boundary, $\partial \Omega$, is surrounded 
by vacuum.
On ${\bf R}^3 \setminus \Omega$ the current 
is ${\bf j}=0$ and therefore in the
steady state 
 the local magnetic field ${\bf H}$ satisfies the Maxwell equations
\begin{mathletters}
\label{ou} 
\begin{equation}
\nabla \cdot {\bf H}=0 
\label{oua}
\end{equation} 
\begin{equation}
 \nabla \times {\bf H}=0. 
\label{oub}
\end{equation} 
\end{mathletters}
The problem reduces to that of finding a magnetic scalar potential  
$\Phi$ that satisfies
\begin{mathletters}
\label{out}
\begin{equation}
{\bf H}=-\nabla \Phi  
\label{outa}
\end{equation}
\begin{equation}
\nabla^2 \Phi =0. 
\label{outb}
\end{equation}
\end{mathletters}
On $\Omega$ the relevant fields are ${\bf H}$, the
superconducting current ${\bf j}$, and  the ``superfluid
velocity field'' ${\bf v_s}$ [\onlinecite{orlando}] defined as:
\begin{equation}
{\bf v_s}=\frac{\nabla \chi}{2} + \frac{e}{c} {\bf A},
\label{vdef}
\end{equation}
where $\chi$ is the phase of the superconducting order parameter, ${\bf A}$
the vector potential, and $e$ the
proton charge, (with $\hbar=k_B=1$). The field  ${\bf v_s}$ is 
conventionally defined as above, and actually has units of momentum.
The relation between ${\bf v_s}$ and ${\bf H}$ is given by  the 
second London equation [\onlinecite{orlando}]:
\begin{equation}
\nabla \times {\bf v_s}=\frac{e}{c}{\bf H}. 
\label{londoneq}
\end{equation}
In the steady state the appropriate Maxwell equation is Amp\`ere's law,
\begin{equation}
\nabla\times {\bf H}=\frac {4 \pi}{c}{\bf j},
\label{ampere}
\end{equation}
substituting Eq. (\ref{londoneq}) into (\ref{ampere}) we obtain:
\begin{equation}
\nabla\times\nabla\times{\bf v_s}=\frac{4 \pi e}{c^2}{\bf j(v_s)}
\label{maxlon}
\end{equation}
which is the general equation that will be investigated in this article.
It must be supplemented by a relation, ${\bf j(v_s)}$, which will be
discussed in \ref{jvv},  to be substituted in its right hand side (RHS).
Eq. (\ref{maxlon}) must then be solved together with
Eq. (\ref{out}) and the proper boundary conditions.

The required boundary conditions are the following: first at infinity
one must have,
\begin{equation}
-\nabla \Phi ={\bf H_a}. 
\label{bc}
\end{equation}
Second, deep inside the
sample all fields must vanish. Third,  ${\bf H}$ must
be continuous [\onlinecite{london,degennes,reitz}] on the boundary      
$\partial \Omega$. 
Finally, the currents are  confined to the superconducting material,
${\bf j \cdot n=0}$ in $\partial \Omega$, ${\bf n}$ is the unit normal 
pointing outwards.
The first boundary condition (at an open 
boundary over an infinite domain) can be satisfied by construction of the
solution in ${\bf R}^3 
\setminus \Omega$. The remaining 
boundary conditions have to be implemented in the numerical algorithm.

Because of these boundary conditions we see that, as emphasized in the 
Introduction, this problem can indeed be computationally very demanding.
It involves solving nonlinear differential equations, in principle in 
an unbounded region, but with the relevant fields varying very
rapidly in a small region inside the material. We will see  here
and in the next
Sections how these difficulties can be overcome, for suitable geometries, 
by making use of a numerical implementation of a perturbation
scheme.

\subsection{The Magnetic Moment}
\label{moment}

Let us at this point introduce some considerations that point to the 
eventual way out of these numerical difficulties.
We recall that the quantity of interest here is the magnetic moment 
${\bf m}$. 
  From  the current distribution  ${\bf j}$ in $\Omega$, the magnetic
moment can be obtained by volume integration [\onlinecite{jackson}]:
\begin{equation}
{\bf m}=\frac{1}{2 c}\int_{\Omega} d{\Omega} \: {\bf r"}\times {\bf 
j(r")},
\label{magmom}
\end{equation}  
where {\bf r"} is the  position vector for a point in the region $\Omega$.

An important consequence of nonlinear Maxwell-London
electrodynamics and unconventional pairing states is that ${\bf m}$
need not be aligned with the applied field ${\bf H_a}$ even if the latter
is applied along a direction of geometrical symmetry, (along a principal
axis of the demagnetization tensor [\onlinecite{landau}] of the body). 
For simplicity, we will restrict ourselves to this case here, although 
it is straightforward to add the complications arising from a 
non-diagonal demagnetization tensor.

Let us introduce
here, for a typical macroscopic experimental HTSC sample, the ratio
between some length $d$ characterizing its size, and the characteristic
value of the London penetration depth, which we shall denote by $\lambda$.
In a macroscopic sample the ratio $\epsilon=\lambda/d$ is a very small
quantity. This small ratio is essentially what will be used in this 
work to
develop a perturbation method to  compute the magnetic moment 
to first order in $\epsilon$. The starting point of this 
 procedure is the existence of a solution in the limit 
$\epsilon \rightarrow 0$ (when all the nonlinear effects vanish),
which corresponds to imposing  trivial Neumann boundary conditions on 
$\partial \Omega$. For a suitable choice of $\Omega$, such as an ellipsoid,
the form of this  solution may be found analytically. The perturbation method  
may also  be applied, as we shall see, in certain cases when only
 a numerical solution in the small $\epsilon$ limit is available.

The components of ${\bf m}$ parallel and perpendicular to ${\bf H_a}$
applied along a direction of symmetry, can 
generically be written for $\epsilon \ll 1$ in the form
\begin{mathletters}
\label{mmm}
\begin{equation}
 m_{\|}=m_0(1-\alpha_{\|} \:\epsilon+{\it O}(\epsilon^2)) 
\label{mmma}
\end{equation}
\begin{equation}
 m_{\bot}=m_0(\alpha_{\bot} \:\epsilon+{\it O}(\epsilon^2)), 
\label{mmmb}
\end{equation}
\end{mathletters}
where $m_0$  denotes the longitudinal
magnetic moment in the limit $\lambda=0$, (and therefore 
$\epsilon=0$), which 
is proportional to $H_a$. It depends only on the geometry of $\Omega$
and therefore contains no physical information. 
For  $\Omega$ in the shape of an 
ellipsoid, values are given in $\;$                                         
 [\onlinecite{landau}]. 
For finite $\lambda$ there is a reduction, linear in the field
to leading order,
in the absolute value
of $m_{\|}$. This reduction is 
due to current penetration in the material and it  implies 
a positive constant $\alpha_{\|}$.
For a very few simple geometrical shapes and linear,
isotropic, Maxwell-London equations,
values of  $\alpha_{\|}$ are given in textbooks 
[\onlinecite{london,fetwal}].
When nonlinear effects are included, they  contribute a correction
to $\alpha_\|$ linear in the field, but their most conspicuous
effect is the appearance, in general, of nonvanishing values of
$\alpha_\bot$, also proportional to the  
field [\onlinecite{ys,zv}].
It is for this reason that  the transverse component is the physical    
 quantity of interest.

We next derive an alternative expression for {\bf m} that helps to fully 
exploit the  existence of the small parameter $\epsilon$. 
Using Eqs. (\ref{ou}), (\ref{ampere}) 
and  formulas for vector calculus [\onlinecite{jackson2}]
one can transform the 
the quantity in the integrand of (\ref{magmom}):
\begin{equation}
{\bf r} \times (\nabla \times {\bf H})=\nabla ({\bf r} \cdot {\bf H})-
({\bf r} \cdot \nabla) {\bf H}- {\bf H}
\label{m1}
\end{equation}
\begin{equation}
({\bf r} \cdot \nabla) {\bf H}=-\nabla \times ({\bf r} \times {\bf H}) 
-2 {\bf H}.
\label{m2}
\end{equation}
After substitution of Eq. (\ref {m2}) into Eq. (\ref {m1}),
\begin{equation}
{\bf r} \times (\nabla \times {\bf H})=\nabla ({\bf r} \cdot {\bf H})
+\nabla \times ({\bf r} \times {\bf H}) + {\bf H},
\label{m3}
\end{equation}
integration over $\Omega$ and use of Gauss' theorem yields:
\begin{eqnarray}
{\bf m}=\frac{1}{8 \pi}\int_{\partial \Omega} dS 
\:[ {\bf n} \: ({\bf r"}\cdot {\bf H})
+\: {\bf n} \times ({\bf r"}\times {\bf H})] 
+\frac{1}{8 \pi} \int_{\Omega} d \Omega \: {\bf H}\equiv {\bf m_1}+{\bf 
m_2},
\label{m4}
\end{eqnarray} 
 where {\bf r"} is the position vector for a point on  $\partial 
\Omega$.  
The notation ${\bf m_1}$, ${\bf m_2}$ refers to the two terms in the
middle portion of Eq. (\ref{m4}). 
If we recall 
Eq. (\ref {londoneq}) and use an alternative form of Gauss' theorem 
we can also rewrite ${\bf m_2}$ as a surface integral:     
\begin{equation}
{\bf m_1}=\frac{1}{8 \pi}\int_{\partial \Omega} dS 
\:[ {\bf n} \: ({\bf r"}\cdot {\bf H})
+\: {\bf n} \times ({\bf r"}\times {\bf H})], \: \: \: \:
{\bf m_2}=\frac{c}{8 \pi e}\int_{\partial \Omega} dS \: {\bf n}         
 \times {\bf v_s}.
\label{magmoms}
\end{equation}         
The terms ${\bf m_1}$ and ${\bf m_2}$ 
are of different order in $\epsilon$ and  the latter 
is small, i.e. of $O(\epsilon m_0)$. This follows from the volume
expression for ${\bf m_2}$, as seen in Eq. (\ref{m4}): since ${\bf H}$ is
confined to a ``skin'' layer of thickness $\lambda$ from the surface,
we see at once that ${\bf m_2}$ will be of order
$\lambda$ times the applied field, that is, of order $\epsilon m_0$, thus
explicitly vanishing in the zero penetration limit. Alternatively, from
Eq. (\ref{londoneq}) one sees that ${\bf v_s}$ scales as $\lambda$ and
then the same result follows from Eq. (\ref{magmoms}). Specifically if the 
equations and
boundary conditions require the field to decay exponentially far from the
surface (up to polynomial corrections), then after decomposing the
volume integral into surface and normal components we have
\begin{equation}
m_2=\frac{1}{8 \pi} \int_{S'} d S' \int_{0}^{w_{max}} d w
\sum_i (H_{max})_i e^{-\frac{w}{\lambda _i}} \leq \frac{3}{8 \pi} 
{\cal S} max(\lambda _i (H_{max})_i). 
\label{estimate}
\end{equation}
This expression is not proportional to the volume of 
$\Omega$, V, as is the case for $m_0$, but rather to $\lambda 
{\cal S}\: \sim \: O(\epsilon V)$ where 
${\cal S}$ is the surface area of $\partial \Omega$. 
As a result $m_2$ is $O(\epsilon m_0)$.

\subsection{The relation between j and ${\bf v_s}$}
\label{jvv}

We now return to the pending question of the equation relating the fields  
${\bf j}$ and ${\bf v}_s$ needed to 
supplement (\ref{maxlon}). This is given by the usual two-fluid phenomenology
 [\onlinecite{ys,zv}]:
\begin{equation}
{\bf j(v_s)}=-eN_f\int_{FS} d^2s \: n(s) {\bf v}_f [({\bf v}_f \cdot {\bf v_s})
  +2\int^{\infty}_0 d\zeta \: f(E(\zeta)+{\bf v}_f\cdot {\bf v_s})]
\label{nonlinjv}
\end{equation}
where  $N_f$ is the total density of states at the Fermi level,
$n(s)$ is the density of states at point $s$ at the Fermi surface (FS),
normalized to unity, ${\bf v}_f(s)$ is the $s$-dependent Fermi 
velocity, $f$ is the Fermi function,
 with $E(\zeta)=(\zeta^2+\left| \Delta(s) \right|^2)^{1/2}$, 
T the absolute temperature and $\Delta(s)$ is the 
superconducting order (gap) parameter.

In the  simplest  approximation (which we will call the linear case), one  
considers  only the linear terms in the
expression for ${\bf j(v_s)}$, that is, one expands the right side        
of (\ref{nonlinjv}) and writes:
\begin{equation}
{\bf j}=-\frac{c^2}{4 \pi e} \tilde{\Lambda}^{-1}{\bf v_s}
\label{linear}
\end{equation}
where $\tilde{\Lambda}$ is the penetration depth tensor.  In the special 
case of an isotropic superconductor, 
and  {\it only} in this case, 
the fields ${\bf H}$, ${\bf j}$ and  ${\bf v_s}$ all satisfy the vector 
Helmholtz equation. But HTSC's are in general highly anisotropic, layered 
structures with penetration depths much smaller [\onlinecite{plakida}] in the 
layers ($a-b$ planes) than in the 
direction perpendicular to them (along the $c$-axis), so that the  
isotropic limit does not apply. The components
of $\tilde{\Lambda}$ in its diagonal representation are the square of
the London penetration depths, $\lambda_a$,  $\lambda_b$ and  $\lambda_c$ 
and experimentally $\lambda_c$ $\gg$ $\lambda_a, \; \lambda_b$. In the  
numerical examples discussed here we will, for simplicity, neglect the 
comparatively small in-plane anisotropy and use the notation  
$\lambda_{ab}$ for the average of  $\lambda_a$ and  $\lambda_b$.

In the anisotropic case Eqs. (\ref{mmm}) are 
still valid with $\epsilon$ expressed in terms of
an effective penetration depth, primarily determined by
whichever component $\lambda_i$ plays 
the dominant role in the current decay.
For example, for the geometry  considered 
in the next Section, the relevant quantity 
is  the penetration depth in the crystallographic
$a-b$ plane.

In the problem of interest here one must consider,
instead of Eq. (\ref{linear}), the nonlinear terms arising from the full
relation (Eq. (\ref{nonlinjv}))
between ${\bf j}$, and ${\bf v_s}$ and substitute this in the RHS of Eq. 
(\ref{maxlon}). After suitable  assumptions
for the FS and other physical quantities are introduced,
this can be done by performing the FS integrals numerically,
 [\onlinecite{sv}] or, in the low temperature limit, analytically 
[\onlinecite{ys,zv}].
Inclusion of these nonlinear terms is crucial, because,
as  mentioned, the physically important angular
dependence of the transverse magnetic moment arises precisely from
these nonlinear effects. The actual expressions for ${\bf j(v_s)}$ used here are
taken from  [\onlinecite{zv}] and are quoted in Appendix A.
       
\section{Analytical Considerations: Oblate Spheroidal Geometry}
\label{geo}

Experimental samples [\onlinecite{bgg}] in which magnetic measurements are 
performed are in the
shape of a flat ``disk'' with rounded edges, and the axis of revolution
along the crystallographic $c$ axis of the crystal. This is well approximated
by taking $\Omega$
to be a flat ellipsoid 
of revolution (an oblate spheroid). 
 For an oblate spheroid it is possible to find an analytic 
expression for the general form of the potential $\Phi$. 
Therefore, our ideas can be implemented in this geometry in terms of 
analytic expressions in ${\bf R}^3 \setminus \Omega$.

The potential $\Phi$, for $\epsilon =0$, satisfies trivial Neumann boundary
conditions on 
$\partial \Omega$, and the solution  
contains a single parameter which is simply related to $m_{\|}$.  
When the penetration depth
is finite, the longitudinal moment does change, but its correct
value can in principle
be determined from the boundary conditions and the solution
inside, either through an iteration process as described in the next Section,
or, much more efficiently, through the perturbation method we shall
develop.

The fundamental equation (\ref{maxlon}) is not separable in spheroidal
coordinates. (Even the vector Helmholtz equation is not).
Still, it is desirable to employ these coordinates as 
 $\partial \Omega$ is then described by a single 
parameter and this significantly simplifies the process of numerically 
 fulfilling the  boundary conditions. The
simple implementation and discretization of the boundary conditions on
$\partial \Omega$ yields higher accuracy where it is most needed,
since boundary grid points lie on the interface.

We denote the major and minor semiaxes of the spheroid by $A$ and $C$ 
respectively,  and we have $A >> C$ for
actual samples. We take
(see Fig. \ref{fig1}) a coordinate system fixed to the direction
of the magnetic field, with its $z$-axis
parallel to the $c$ crystallographic direction of the superconductor (and
parallel to the $C$ semiaxis of an ellipsoid). 
The field is applied along the $x$-axis, and we picture the experiment
as being performed by rotating the crystal about the $z$-axis and 
measuring the angular dependence of ${\bf m_\bot}$. 
As the crystal is rotated the axes $x-y$ remain fixed in space, and 
should not be confused with the  coordinates, affixed to the
crystal structure, that we shall also use and denote by $x'$, $y'$.     
 We call $\Psi$  the angle between axes $x$ and $x'$.

In the definition we use [\onlinecite{magnus}], the oblate spheroidal   
  coordinates $\xi, \eta, \varphi$, are related to Cartesian
coordinates by the transformation
\begin{mathletters}
\label{coor}
\begin{equation}
        x = f (1+\xi^2)^{1/2}(1-\eta^2)^{1/2} \cos \varphi, \\
\label{coora}
\end{equation}
\begin{equation}
        y = f (1+\xi^2)^{1/2}(1-\eta^2)^{1/2} \sin \varphi, \\
\label{coorb}
\end{equation}
\begin{equation}
        z = f \xi \eta,
\label{coorc}
\end{equation}
\end{mathletters}
where $0 \leq \xi < \infty$, $-1 \leq \eta \leq1$, $0\leq \varphi \leq 
2\pi$,
and $f$ is a focal length scale factor (the
coordinates $\xi$, $\eta$ and $\varphi$ are dimensionless). In Fig. 2 we 
show this coordinate system at a fixed azimuthal angle 
$\varphi=0^\circ$. One can obtain the
relation between  $\xi$, $\eta$, $\varphi$ and
 Cartesian coordinates fixed to the crystal
by replacing $\varphi$ with $\varphi+\Psi$ in (\ref{coor}). For example
$x' = f (1+\xi^2)^{1/2}(1-\eta^2)^{1/2} \cos (\varphi+\Psi)$.
The relation between unit vectors in these and Cartesian
coordinates is
\begin{mathletters}
\label{unit}
\begin{equation}
\hat \xi=\frac{1}{(\xi^2+\eta^2)^{1/2}}
(\xi (1-\eta^2)^{1/2} \cos \varphi \hat x
+\xi (1-\eta^2)^{1/2} \sin \varphi \hat y
+(1+\xi^2)^{1/2} \eta \hat z), 
\label{unita}
\end{equation}
\begin{equation}
\hat \eta=-\frac{1}{(\xi^2+\eta^2)^{1/2}}
((1+\xi^2)^{1/2} \eta \cos \varphi \hat x
+ (1+\xi^2)^{1/2} \eta \sin \varphi \hat y
-  \xi (1-\eta^2)^{1/2} \hat z), 
\label{unitb}
\end{equation}
\begin{equation}
\hat \varphi= -\sin \varphi \hat x + \cos \varphi \hat y 
\label{unitc}
\end{equation}
\end{mathletters}
We see that $\hat \xi= {\bf n}$ is the unit normal pointing outwards.
In the limit  $f \rightarrow  0$, the spheroidal system reduces to 
the
spherical coordinate system. For $f$ finite, the surface $\xi = const$
becomes spherical as $\xi \rightarrow \infty$:
\begin{equation}
f \xi \rightarrow r, \: \:\eta \rightarrow \cos \theta, \: \:
as \: \:\xi \rightarrow
\infty. 
\label{limit}
\end{equation}
In the same limit $\hat \xi \rightarrow \hat r$ and $\hat \eta
\rightarrow \: -\hat \theta$,
where $r$ and $\theta$ are spherical coordinates. Various quantities in 
oblate spheroidal coordinates are given in Appendix B.

To construct the 
general form of the solution for the fields in the region
${\bf R}^3 \setminus \Omega$ we employ an electrostatic analogy         
[\onlinecite{landau}]. 
The current distribution is localized within $\Omega$ and the magnetic
potential in the exterior region
can be written as an expansion in the appropriate set of orthogonal functions
which in this case are the spheroidal harmonics 
[\onlinecite{magnus,dover}]
(a generalization of spherical harmonics), characterized by angular and
azimuthal indices $l$ and $m$. Thus we write
\begin{equation}
\Phi = \Phi_a+ \sum_{l=0} ^{\infty}\sum_{m=0} ^l 
(A_l^m \cos (m\varphi)+ A_{l \bot}^m \sin (m \varphi))
 f Q_l^m(i\xi))  P_l^m(\eta) \: \: \: \: \: \xi \ge \xi_0,
\label{figen}
\end{equation}
where $P^{m}_{l}$ and $Q^{m}_{l}$ are the associated Legendre functions of 
the first and second kind respectively, and 
$\Phi_a$ is the potential due to the applied field.
The condition that $-{\bf \nabla} \Phi$ 
$\rightarrow$ $H_a \hat{x}$ when $\xi \rightarrow \infty$  yields 
\begin{equation}
\Phi_a=-H_a f P^1_1(i\xi) P^1_1(\eta)\cos\varphi. 
\label{fa}
\end{equation} 
Since for $\xi \rightarrow \infty \:$ $Q_l^m(i\xi) \rightarrow 0 \:$ 
$\forall$ $m$, $l$ this fulfills the boundary condition at infinity. 
The remaining terms in Eq. (\ref{figen}) are due to the presence of the
superconductor. The coefficients $A_{1\|}^1$ and $A_{1 \bot}^1$, for example, 
multiply terms that give rise to magnetic fields associated
with dipole moments  along ${\bf 
H_a}$ and perpendicular to it, respectively. 
In the limit $\epsilon=0$  Eq. (\ref{figen}) simplifies 
[\onlinecite{landau}] and
the exact solution for $\Phi$ is
\begin{equation}
\Phi=\Phi_a+ A_{1\|} f Q^1_1(i\xi) P^1_1(\eta) \cos\varphi.  \label{phiout} 
\end{equation}
The parameter $A_{1\|} \equiv A^1_{1\|}$ (for brevity we omit the 
upper index $m=1$)
is determined from the boundary condition $\partial_{\xi} \Phi=0$ at 
$\xi=\xi_0$, the surface of the spheroid:
\begin{equation}
A_{1\|}(\epsilon=0)=\frac{- 
H_a\xi_0}{1+1/(1+\xi_0)-\xi_0\arctan(1/\xi_0)}.
\label{a1}
\end{equation}
$A_{1\|}$ is always proportional to $m_{\|}$, the 
longitudinal magnetic moment. For an ellipsoid, we show in Appendix A that
\begin{equation}
m_{\|}=\frac{2}{3}f^3A_{1\|}.
\label{longmag}
\end{equation}

It is instructive to verify that in the linear case  the solution on
$\Omega$ leads to vanishing $m_{\bot}$. From  the
linear relation between ${\bf j}$ and ${\bf v_s}$ in
Eq. (\ref{linear}) and the anisotropy in $\tilde{\Lambda}$ as discussed 
in \ref{jvv}, the azimuthal dependence of ${\bf v_s}$ and ${\bf 
j}$ is identical. We can find
 the $\varphi$-dependence of the magnetic moment from Eq. 
(\ref{magmom}) and the appropriate element of integration given 
by Eq. (\ref{domega}). The $\varphi$ variable is separable. 
Consequently, the  $\int ^{2 \pi}_0 d \varphi$ integration can be  performed 
analytically and yields  $m_{\bot}=0$. Thus, any transverse
component arises from the nonlinear terms. Their origin can be seen
as follows:
the superfluid velocity field has, in the presence of nonlinearities,
an azimuthal dependence different from  that in
the linear case. This implies that the
$\varphi$ variable is no longer separable.
The nonlinear terms in (\ref{jnl}) lead to higher
harmonics for the $\varphi$ dependence of ${\bf v_s}$ on $\Omega$. 
Then, it follows from Eq. (\ref{londoneq}) that there is a 
small, but nonvanishing transverse field, $H_{\bot}$,
with a different azimuthal dependence from that in the linear case, which can 
contribute to $m_{\bot}$.
Therefore the nonlinear response of a superconductor is
responsible for $H_{\bot}$ and, as a result, for  $m_{\bot}$. In ${\bf R^3}
\setminus \Omega$, the part of $H_{\bot}$ from which  $m_{\bot}$ arises 
can be described by a transverse dipole, i.e. a potential of the form of the  
 last term in Eq. (\ref{phiout}), rotated by $90^\circ$:
\begin{equation}
\Phi_{\bot}=A_{1\bot} f Q^1_1(i\xi) P^1_1(\eta) \sin\varphi 
\label{phiout2}
\end{equation}
where in full analogy to the longitudinal case in Eq. (\ref{longmag})
\begin{equation}
m_\bot=\frac{2}{3}f^3 A_{1\bot}. \label{perpmag}
\end{equation}
The potential $\Phi_{\bot}$ is the only contribution to $m_{\bot}$ from
the general expression given by Eq. (\ref{figen}). This term is small,
$A_{1\bot} \ll A_{1\|}$ (recall the discussion about $m_{\bot} / m_{\|}$
from subsection \ref{moment}) and does not
exist when the penetration depths vanish, since the nonlinear effects 
are absent unless the field can penetrate the sample.
Higher order multipole terms $(l>1)$, as in (\ref{figen}) do not 
contribute to the magnetic moment as can be seen from symmetry 
considerations or by explicit calculation using the orthogonality  of   
$P^m_l(\eta)$.

It is useful to recall explicitly the connection of the $A_l^m$ with the
coefficients in a standard multipole 
expansion. In the spherical limit given by Eq. (\ref{limit}) 
these coefficients represent ordinary spherical multipoles. 
For large enough distances from $\Omega$ the asymptotic form of the
$l=1$ term is always a pure dipole.
The spherical limit of Eq. (\ref{phiout}) is
\begin{equation}
\Phi=-H_a r \sin \theta \cos \varphi +
\frac{m_0}{r^2} \sin \theta \cos \varphi=-H_a x+\frac{m_0 x}{r^3}, \label{fis}
\end{equation}
with
\begin{equation}
m_0=-\frac{1}{2} H_a a^3 
\label{mos}
\end{equation}
where $a$ is the sphere radius and we see once more that 
$A_{1\|}(\epsilon=0)$ is proportional to $m_0$.

For finite $\xi$ (not necessarily in the spherical 
limit), the magnetic moment of a spheroid is again 
obtained only  from terms with $l=1$ 
in the Eq. (\ref{figen}). 
The $l=1$ terms, however,  
have spheroidal symmetry and  their radial and angular
dependence is not identical to $\frac{1}{r^2} \sin \theta$ as for a pure
dipole in Eq. (\ref{fis}), they have an admixture of higher
spherical multipoles.

At finite $\tilde{\Lambda}$
the value for 
$A_{1\|}$ (proportional to $m_{\|}$ from (\ref{longmag}))
 will be slightly different from $A_{1\|}(\epsilon=0)$, reflecting the   
difference between $m_{\|}$ and $m_0$ given by Eq. (\ref{mmma}). The 
potential will also acquire additional terms as seen in Eq. 
(\ref{figen}). The longitudinal
terms with $l\geq 2$ and all of the transverse terms
vanish at $\tilde{\Lambda}=0$, that is, they are of
higher order in $\epsilon$. They also vanish 
in the spherical case but only if $\tilde{\Lambda}$ is
isotropic and the relation ${\bf j(v_s)}$ is
purely linear.  If, in addition, one includes the nonlinear terms in
${\bf j(v_s)}$, so that the full relation (\ref{figen}) applies, then 
the transverse dipole term appears even in a spherical geometry. 

\section{Computation of the Magnetic Moment}
\label{comp}
\subsection{General Considerations on Iteration Procedure}
\label{conv}
The considerations from the previous section point to a method 
for obtaining a complete
solution for ${\bf m}$. This method, although it should work in principle, 
  would naively lead to the need for an iteration which is in
practice too cumbersome, and which we discuss here as motivation for
introducing the perturbation method obviating the need for it.
The important question here is that of calculating the magnetic dipole 
moment. In other boundary value problems to which our method can be 
extended, one must similarly determine the lowest nonvanishing multipole 
moments. 

To obtain a solution to  Eq. (\ref{maxlon}) in the general case when 
the relation between ${\bf j}$ and ${\bf v_s}$ is nonlinear, given by
Eq. (\ref{nonlinjv}) (see also (\ref{jnl})), one can in principle use the 
following iteration method.
As a first step, one can solve these equations in $\Omega$
with boundary conditions on 
$\partial \Omega$ corresponding to
the limit $\epsilon=0$. That can be done by  setting
$A_{1\|}=A_{1\|}(\epsilon=0)$ (and all the other $A_{l \: \|,\: 
\bot}^m\equiv0$), requiring continuity of the components
$H_{\eta}$ and $H_{\varphi}$ at $\xi=\xi_0$, and also enforcing
the condition on $\partial \Omega$, ${\bf j} \cdot {\bf n}=0$. 
These boundary conditions for the magnetic field are not exactly the desired  
 ones, since continuity of the $\xi$ component cannot be demanded
because the external field has been specified so that its $\xi$
component vanishes at the boundary. 
The magnetic field outside the sample is  simply
obtained from Eq. (\ref{outa}) with $\Phi$ given by (\ref{phiout}). 
Inside one employs Eq. (\ref{londoneq}) to get the magnetic field from 
${\bf v_s}$. 

 From the numerical solution in $\Omega$, obtained using the 
overrelaxation method discussed in [\onlinecite{rec,smith}] we can compute
the magnetic moment by using (\ref{magmom}), and the fields from the  
above approximate solution. 
The computed value, ${\bf m^{(1)}}$, the superscript $(1)$ indicating 
the order of iteration, will be in general different from 
$m_0 \hat {x}$: the computed magnetic moment is not the 
input value $m_0$. Hence, the actual
problem has not been solved: the solution is not self-consistent.
One can then imagine obtaining the correct solution
from the following iteration process:
Denote by $(A_{1\|})^{(1)}$, $(A_{1\bot})^{(1)}$ the values of these quantities
obtained from Eqs. (\ref{longmag}), Eq. (\ref{perpmag}) and the appropriate 
components of ${\bf m^{(1)}}$. Then use $(A_{1\|})^{(1)}$, 
$(A_{1\bot})^{(1)}$ in the computation of the exterior field, and use 
again this exterior field to solve Eq. (\ref{maxlon}), repeating the 
procedure described in the previous paragraph.
The second iteration 
yields e.g. $(A_{1\|})^{(2)}$, $(A_{1\bot})^{(2)}$ which in general would differ
from  $(A_{1\|})^{(1)}$ and $(A_{1\bot})^{(1)}$.  
Repeated iterations would give  sequences 
 $(A_{1\|})^{(1)}$,  $(A_{1\|})^{(2)}$,  $(A_{1\|})^{(3)}$,... 
 and $(A_{1\bot})^{(1)}$, $(A_{1\bot})^{(2)}$, $(A_{1\bot})^{(3)}$,...
 converging to the desired values of  $A_{1\|}$, $A_{1\bot}$,
 when the moment 
generated by the computed currents equals the input value.  At that 
point, the magnetic moment will be known. The higher order $A's$ will 
not necessarily be known, but they do not contribute to ${\bf m}$.
In practice such a procedure could be implemented first for the larger 
longitudinal component and then for the smaller transverse component.

\subsection{Perturbation Method To Compute  m} 
\label{per}

The procedure described in \ref{conv}
 may be a lengthy and expensive process and it is for 
 this reason that we develop, in this subsection, a procedure to bypass 
it. We compute $m_{\|}$ (accurate to first order in $\epsilon$)
in a single iteration step, that is, a single pass through solving
the equations inside the body as described above.

The surface integral (\ref{m4}), rather than (\ref{magmom}) is the 
expression for ${\bf m}$ that is convenient for our purpose since as 
seen in subsection \ref{moment}, it divides ${\bf m}$ into two terms, 
${\bf m_1}$ and ${\bf m_2}$, of different order in $\epsilon$. 

It follows from the considerations of Section \ref{maxw}, 
and it is the basis of our perturbation method, 
that to obtain ${\bf m}$ correctly to $O(\epsilon m_0)$
it is sufficient to compute ${\bf m_2}$ from fields (i.e. ${\bf H}$ or
${\bf v_s}$) accurate only to zeroth order. 
This is, as explained there, because a factor of $\epsilon$ explicitly 
scales out of the expression for ${\bf m_2}$. Now, since the internal 
fields obtained by solving  Eq. (\ref{maxlon}) at the first iteration 
level (as described in the previous subsection, i.e., with $\epsilon=0$ 
boundary conditions) are already accurate to the zeroth order, one 
iteration is sufficient to evaluate ${\bf m_2}$ at desired accuracy. 
The problem reduces, therefore, to that of correctly including the 
contribution ${\bf m_1}$ to first order in $\epsilon$.

To illustrate how this is done, let us recall the general form of the 
analytic solution for the spheroid. As seen in Section \ref{geo} 
(Eq. (\ref{figen})) it has the form of a multipole  expansion with      
undetermined coefficients. Since the exact field ${\bf H}$ is continuous  
on $\partial \Omega$ we can insert this general form 
in the expression (\ref{m4}) for ${\bf m_1}$. Only 
the $l=1$  terms, by virtue of (\ref{longmag}) and (\ref{perpmag}), 
 contribute to ${\bf m}$, so that  ${\bf m_1}$ can be evaluated 
in terms of the unknown ${\bf m}$.
Thus we have
\begin{equation}
{\bf m}={\bf m_1 (m) }+{\bar{{\bf m}_2}} +O(\epsilon ^2 m_0),
\label{12}
\end{equation}
where we emphasize that  ${\bf m_1}$ depends on the unknown $l=1$ parameters, 
and where we introduce the overbar notation to denote quantities evaluated  
from the zero penetration limit external fields.  Since 
terms of $O(\epsilon ^2 m_0)$ can be neglected,  it is possible to 
interchange ${\bf m_2}$ and ${\bar{{\bf m}_2}}$ in the various 
expressions and we have done so. Expression (\ref{12}) is an equation for the 
unknown ${\bf m}$.

To solve it in practice, consider the general form of $\Phi$, Eq. 
(\ref{figen}) which indicates that $\Phi$ (or 
equivalently ${\bf H}$) 
can be separated into two parts: $\Phi=\Phi_a+\Phi_r$, 
due to the applied field and to the 
presence of the superconductor, respectively. We can then write the 
contribution of these parts as
\begin{equation}
{\bf m_1(m)}={\bf m_1}(\Phi_a)+{\bf m_1}(\Phi_r). 
\label{m1divide}
\end{equation}
We now define $p$ by ${\bf m_1}(\Phi_a)\equiv p {\bf m_0}$. Since ${\bf m_0}$  
 and ${\bf m_1}(\Phi_a)$ are now known, one can determine the constant $p$  
which  depends on the shape of $\Omega$, i.e.  on the eccentricity.
In the limit $\epsilon=0$, ${\bf m_2}=0$ and from Eq. (\ref{m1divide}) 
we have the identity
\begin{equation}
{\bf m_0}= {\bf \bar{m}_1}= p {\bf m_0}+ {\bf m_0} (1-p).  
\label{m1zero}
\end{equation}
For $\epsilon \neq 0$, when the solution for $\Phi$ 
and ${\bf H}$ is given 
in terms of multipole expansion with unknown coefficients, ${\bf 
m_1}(\Phi_a)$ remains the same. The only difference in computing ${\bf 
m_1}$ is that the coefficients in the terms arising from $\Phi_r$
are now proportional to the correct, but still to be determined, value
of the magnetic moment ${\bf m}$,
slightly changed from the $\epsilon=0$ case. All the remaining higher 
multipoles $(l >1)$ of $\Phi$, 
as mentioned in \ref{geo}, do not contribute to ${\bf m}$ and we have
\begin{equation} 
{\bf m_1}(\Phi_r)={\bf m} (1-p). 
\label{m1rest}
\end{equation}
Adding this to  ${\bf m_1}(\Phi_a)$ we get   
\begin{equation}
{\bf m_1}={\bf m}-p({\bf m}-{\bf m}_0).
\label{mp}
\end{equation}
We can express Eq. (\ref{magmoms}) using (\ref{mp}) as
\begin{equation}
{\bf m}={\bf m}-p({\bf m}-{\bf m}_0)+{\bf m}_2,
\label{mp2}
\end{equation}
so that we have the solution for ${\bf m}$ correct to $O(\epsilon)$,
\begin{equation}
{\bf m}={\bf m}_0+\frac{1}{p} {\bf m}_2\approx 
{\bf m}_0+\frac{1}{p} {\bf \bar{m}}_2,
\label{mwithp}
\end{equation} 
which determines all components of ${\bf m}$.

We illustrate this  method using the textbook example of the 
isotropic, linear superconducting sphere in a uniform 
applied field ${\bf H_a}$, 
along the x-axis. In this case all the fields
in $\Omega$ satisfy the vector Helmholtz 
equation:
\begin{equation}
{\bf \nabla}^2 {\bf F}=\frac{1}{\lambda ^2} {\bf F}, \label{helmholtz}
\end{equation}
where ${\bf F}$ can be ${\bf H, \; j ,\; v_s}$.
On the entire ${\bf R^3} \setminus \Omega$ region, $\Phi_r$ has a pure 
dipole form and ${\bf H}$ is given 
by taking the gradient of Eq. (\ref{fis}):
\begin{mathletters}
\label{hs}
\begin{equation}
H_r=(H_a+\frac{2m}{r^3}) \sin\theta \cos \varphi, 
\label{hsa}
\end{equation}
\begin{equation}
H_\theta=(H_a-\frac{m}{r^3}) \cos \theta cos \varphi, 
\label{hsb}
\end{equation}
\begin{equation}
H_\varphi=(-H_a+\frac{m}{r^3}) sin\varphi, 
\label{hsc}
\end{equation}
\end{mathletters}
where $r$, $\theta$, $\phi$ are spherical coordinates, and $m$ a parameter
to be determined.
This is the general solution for any $\epsilon$ (not just 0), in  the 
field outside; there is only a dipole term in  addition to that 
due to applied field. However, even if there were higher spherical 
multipoles in the general solution, that would not affect the evaluation 
of $m_1$, since their contribution to ${\bf m}$ would vanish identically 
by symmetry. In the limit where the  
current does not penetrate into the superconductor, the magnetic moment is 
  given by $m_0=-\frac{1}{2}a^3 H_a$ (recall (\ref{mos})).
Performing the
elementary integral for ${\bf m}_1$ in Eq. (\ref{magmoms}) we obtain 
\begin{equation}
  {\bf m}_1=\frac{1}{3}{\bf m}_0 + \frac{2}{3} {\bf m}=
{\bf m}-\frac{1}{3} ({\bf m-m}_0),
\label{ms1}
\end{equation}
Comparing with  (\ref{mp}) we identify $p=\frac{1}{3}$ in this case and 
using Eq. (\ref{mwithp})
\begin{equation}
 {\bf m}= {\bf m}_0+ 3 \bar{\bf m}_2. \label{ms}
\end{equation}
To determine the unknown $m$ to $O(\epsilon)$
it remains to compute  $\bar{\bf m}_2$ 
and substitute it in (\ref{ms}).
Using the evaluation  for $\bar{\bf m}_2$, with the boundary conditions 
taken in the $\epsilon=0$,  from Appendix C we get  the perturbation 
result for $m$:
\begin{equation}
m=m_0(1-3 \epsilon),  
\label{exact}
\end{equation}
where $\epsilon = \lambda /a$.
 If we compare this to the expression for $m_{\|}$ given by Eq. 
(\ref{mmma}) we can read off $\alpha_{\|}=3$. This is the 
correct value for a sphere to this order, as given in textbooks 
[\onlinecite{london,reitz}].
Thus, using the perturbation method with 
approximate boundary conditions in the evaluation of $m_2$,
we have obtained the correct value for $m$ to first order in 
$\epsilon$. 
It is also instructive to calculate $m$ analytically,
with internal fields evaluated from (\ref{maxlon}) and $\epsilon=0$ boundary 
conditions, Eq. (\ref{m4}). The calculation would 
 yield $m$ correct only to $O(m_0)$; the value of 
  $\alpha_{\|}$ is not correct; one gets  $\alpha_{\|}=2$ instead of     
  $\alpha_{\|}=3$.

We return now to the oblate spheroid. 
In $\Omega$ we allow the full nonlinear relation between 
${\bf j}$ and ${\bf v_s}$, given by Eq. (\ref{jnl}). 
The magnetic field in 
 ${\bf R}^3 \setminus \Omega$ that contributes to 
the computation of ${\bf m}$ is given in Appendix \ref{field}.
We aim to obtain the appropriate perturbation 
equation (\ref{mwithp}) relating the unknown magnetic moment 
to  ${\bar {\bf m}}_2$, the term computed from the  numerical 
solution in $\Omega$ (using the first step in the iteration procedure, 
described in the previous subsection). 
To calculate the magnetic moment, we proceed as with the 
sphere example. 
We first evaluate the term ${\bf m_1}$ from Eq. (\ref{magmoms}), recalling 
Eqs. (\ref{a1}), (\ref{longmag}), (\ref{perpmag}) and 
${\bf n}={\hat {\xi}}$. The integral, 
 evaluated in spheroidal coordinates using 
(\ref{domega}), is elementary and we give only the result.
The corresponding perturbation equation for ${\bf m}$ is, from 
(\ref{mwithp}):
\begin{equation}
{\bf m}={\bf m}_0+\frac{1}{p(\xi_0)} {\bf m}_2, \label{mel}
\end{equation}
or, writing its components explicitly:
\begin{mathletters}
\label{me}
\begin{equation}
 m_{\|}=m_0+\frac{1}{p(\xi_0)} m_{2\|}, \label{mea}
\end{equation}
\begin{equation}
 m_{\bot}=\frac{1}{p(\xi_0)} m_{2\bot}, \label{meb}
\end{equation}
\end{mathletters}
with
\begin{equation}
p(\xi)=\frac{1}{4}(2+\xi ^2-(1+\xi^2) \xi arctan(1/\xi)), \label{pofe}
\end{equation}
$p(\xi)$ is evaluated at the surface of the ellipsoid $(\xi= 
\xi_0)$. As shown in Appendix \ref{mom}, $\xi_0$ is related to the 
eccentricity of the spheroid. 
In the spherical limit when $\xi \rightarrow \infty$ we 
recover the
spherical result, $p=1/3$,  obtained earlier. Another interesting limit 
is that of a flat disk $(\xi \rightarrow 0)$
where $p(\xi)=1/2$. As discussed previously, 
$m_{\bot}\rightarrow 0$ for $\tilde{\Lambda} \rightarrow 0$.

 The above method applies not only to oblate spheroids, but to 
all geometries for which a general  solution for $\Phi$ in  
 $R^3 \setminus \Omega$ as an expansion in terms of orthogonal functions, 
(only one of them being dipolar at large distances) can be written. 
Furthermore, the method can be extended, with one additional 
assumption, to a situation where the shape of $\Omega$ precludes an 
analytic solution for the outside fields even at $\epsilon=0$, and only 
a numerical solution, $\overline{\Phi}$,  in that limit is available.
We assume that, as in the analytic cases, the coordinate dependence 
of the terms in $\Phi$ which contribute to the dipole ${\bf m}$ remains 
the same, up to a multiplicative constant, for  $\epsilon=0$ and
 $\epsilon\neq0$. This assumption requires that the shape of $\Omega$   
 produces no singularities in the fields. 
 This might be a sufficient condition, but we know of no rigorous proof.

 The magnetic field at large 
distances, $r \gg d$,  has the form given in Eq. 
(\ref{fis}), $m_0$ is the magnetic moment for $\Omega$, and all the 
higher multipoles can be neglected. 
We can again, as above (\ref{m1divide}) separate $\overline{\Phi}=       
\Phi_a \:+ \: \overline{\Phi}_r$, where $\Phi_a$ is the applied field 
contribution.  At $r \gg d$, $\overline{\Phi}_r$ is of
 dipolar form and the value of $m_0$ can be
in principle numerically extracted either by using the left part of 
(\ref{m1zero}) or from the asymptotic form.
For $\epsilon \neq 0$ the potential $\Phi$ has asymptotically the same dipolar 
form as $\overline{\Phi}$,  but the unknown dipolar coefficient ${\bf m}$ 
differs from $m_0 \hat{x}$. One can then proceed as in the analytic 
case. Consider as an illustration, the computation of $m_{\|}$ to 
$O(\epsilon)$. We can implement the method by writing the potential at 
$\epsilon \neq 0$ in the form:
\begin{equation} 
\Phi(\epsilon 
\neq0)=\Phi_a+\frac{m_\|}{m_0}(\overline{\Phi}_r)+\Phi_{nd},
\label{near}
\end{equation}
where $\Phi_{nd}$ is a possible contribution to higher order multipoles 
only. Eq. (\ref{near}) merely expresses our assumptions in mathematical 
form. It can be better understood by recalling the discussion in Section 
\ref{geo} (and above in this Section) in particular the difference 
between $A_{1\|}(\epsilon=0)$ and $A_{1\|}$. 

Since $m_0$ is known, one can use $\Phi_a$ to evaluate ${\bf 
m_1(\Phi_a)}$ (see (\ref{m1divide})) and hence the quantity $p$ through 
$m_1(\Phi_a)=p m_0$.
From the distribution of ${\bf v_s}$ on $\partial \Omega$ we compute    
  ${\bar{m}_2}$ and  Eq. (\ref{mwithp}) gives the desired $m_\|$.

We see therefore that our method has considerable generality and our 
results can be summarized in terms of the following 
${\bf Theorem}$, the validity of which  follows from the analysis in   
 this section and the decomposition of ${\bf m}$ in \ref{moment}.

Let us assume that: \\
a) There exist a small parameter $\epsilon$ $\ll$ 1 and we consider Eq. 
(\ref{maxlon}) in $\Omega$ that allows a sufficiently accurate solution 
 in ${\bf R}^3 \setminus \Omega$
 for (\ref{outb}) with trivial Neumann boundary conditions 
 on $\partial \Omega$, and at infinity $-\nabla \Phi= {\bf H_a}$, which 
satisfies the assumption discussed in connection with 
(\ref{near}). \\
b) ${\bf H}$ in  the interior of  $\Omega$ 
decreases  with the distance from $\partial \Omega$
not slower than  exponential dependence given by a 
characteristic length $\ll$ typical size of $\Omega$ \\
Then the following statements are true:

\noindent 1.It is possible to write 
${\bf m}={\bf m}_1+{\bf m}_2$ as given by (\ref{magmoms}) 
where $m_1$ is O$(m_0)$ and 
$m_2$ is O($\epsilon m_0$). \\
\noindent 2. To obtain ${\bf m}$ from Eq. (\ref{mwithp}) 
accurate  to O($\epsilon m_0$)
 it is sufficient to 
calculate the leading contribution to the term ${\bf m}_2$, the error in determining 
${\bf m}$ being of O($\epsilon ^2 m_0$).

This perturbation method can be applied outside the field of 
superconductivity. For example, it
is well known [\onlinecite{landau}] that an ordinary conductor in 
a high frequency, harmonic applied magnetic field (the frequency should  
satisfy quasi-static condition $\omega \ll c/d$) 
behaves like a superconductor in a constant field. It is then possible 
to identify the small parameter, $\epsilon \ll 1$ as the ratio of skin depth, 
$\delta_s$, the typical length scale for field penetration in the 
conductor and the characteristic geometrical dimension, $d$. 
The computation of the magnetic field distribution is then achieved by 
solving the corresponding steady-state problem for a superconductor of 
the same shape, and ${\bf m}$ can be obtained using Eq. (\ref{mwithp}).

\section{Numerical Considerations}
\label{inside}
\subsection{Dimensionless form of Equations on $\Omega$}
\label{dform}
In performing the calculations and describing the results, it is
convenient to introduce dimensionless quantities. 
 We recall that $\Omega$ is a flat spheroid and with 
the magnetic field applied in the $x-y$ ($a-b$) plane, most of the 
current will flow parallel to the $a-b$ plane and its decay will be 
governed by $\lambda_{ab}$. It is therefore convenient to measure
 the length 
in  the units of $\lambda_{ab}$. 
We then define dimensionless fields ${\bf V}$, ${\bf J}$, and ${\bf 
{\cal H}}$:
\begin{equation}
{\bf V}=\frac{{\bf v_s}}{v_c},\qquad {\bf {\cal H}}=\frac{{\bf H}}{H_0},\qquad 
{\bf J}=\frac{c H_0}{4\pi \lambda_{ab}} \, {\bf j},
\label{dimless}
\end{equation}
where $v_c=\Delta_0/v_f$ is the critical velocity (discussed in 
Appendix A), $\Delta_0$ is the amplitude of the 
order parameter, defined in Appendix A,
 and  we have introduced 
a characteristic magnetic field $H_0$ as 
\begin{equation}
H_0=\frac {\phi_0}{\pi ^2 \lambda_{ab} \xi_{ab}},
\label{h0}
\end{equation}
where $\phi_0=\pi c \hbar /e$ is the flux quantum and 
$\xi_{ab}=v_f/\pi \Delta_0$ is the 
in-plane superconducting coherence length (not to be confused with 
$\xi$, the spheroidal coordinate). 
The definition  
(\ref{h0}) involves precisely the same numerical factors as that
used in  [\onlinecite{sv}, \onlinecite{zv}].
The required equations are easily rewritten in terms of these quantities.
Equation (\ref{maxlon}), using the relation between 
${\bf j}$ and ${\bf v_s}$ given by (\ref{jnl}), then becomes
\begin{mathletters}
\label{maxlon2}
\begin{equation}
( \nabla\times\nabla\times {\bf V})_{x',y'} =- V_{x',y'} 
(1-t_1\left|V_{x',y'}
\right|) \equiv - V_{x',y'}+N_{x',y'},
\label{maxlon2a}
\end {equation}
\begin{equation}
(\delta \nabla\times\nabla\times {\bf V})_{z} =
-V_z(1-t_2\frac{V_{x'}^2+V_{y'}^2}
{\left|V_{x'}\right|+\left|V_{y'}\right|}) \equiv - V_{z}+N_{z},
\label{maxlon2b}
\end{equation}
\end{mathletters}
where $\delta=(\lambda_c/\lambda_{ab})^2=m_c/m_{ab}$ and we define 
$N_{x'}, \: N_{y'}, \: N_{z}$ as the terms nonlinear in the velocity. 
 The equations are written in terms of the primed coordinates
and the derivatives are with respect to the dimensionless length measured in 
 units of $\lambda_{ab}$.

Before discretizing Eqs. (\ref{maxlon2}) we
 transform them to oblate spheroidal coordinates. We start by
writing these equations in the unprimed $(x, y, z)$ coordinate system where,
we recall, the $x$-axis lies along the applied field. The linear part of
the equations looks identical in primed or unprimed coordinates and we only 
need
to carefully transform $N_{x', y', z}$ to $N_x$, $N_y$ and $N_z$ 
terms nonlinear in the velocity along the unit vectors
$\hat x$, $\hat y$ and   $\hat z$ respectively. We have
\begin {mathletters}
\label{npart}
\begin{equation}
  N_x=t_1 (V_{x'}\left|V_{x'}\right| \cos \Psi
            +V_{y'}\left|V_{y'}\right| \sin \Psi),
\label{nparta}
\end{equation}
\begin{equation}
  N_y=-t_1 (V_{x'}\left|V_{x'}\right| \sin \Psi
            -V_{y'}\left|V_{y'}\right| \cos \Psi),
\label{npartb}
\end{equation}
\begin{equation}
  N_z=t_2 V_z\frac{V_{x'}^2+V_{y'}^2}
{\left|V_{x'}\right|+\left|V_{y'}\right|},
\label{npartc}
\end{equation}
\end{mathletters}
where $V_{x'}=V_x \cos \Psi-V_y \sin \Psi$,
 $V_{y'}=V_x \sin \Psi+V_y \cos \Psi$ and $V_{z'}=V_z$. Or, if we
express the components of velocity in spheroidal coordinates:
\begin{mathletters}
\label{vesp}
\begin{equation}
V_{x'}=\cos (\varphi +\Psi) (a V_{\xi}+d V_{\eta})
      -\sin(\varphi+ \Psi) V_{\varphi}, \label{vespa}
\end{equation}

\begin{equation}
V_{y'}=\sin (\varphi +\Psi) (a V_{\xi}+d V_{\eta})
      +\cos (\varphi+ \Psi) V_{\varphi}, \label{vespb}
\end{equation}

\begin{equation}
V_{z}=-d V_{\xi}+a V_{\eta}, \label{vespc}
\end{equation}
\end{mathletters}
where $a=\frac{\xi(1-\eta^2)^{1/2}}{\xi^2+\eta^2}$ and
 $d=-\frac{(1+\xi^2)^{1/2} \eta}{\xi^2+\eta^2}$.
We can now write the nonlinear part (from equation
(\ref{maxlon}) and (\ref{npart})) along each spheroidal coordinate.
For example, along $\hat \xi$ we get
\begin{equation}
N_{\xi}=a(\cos \varphi N_x+ \sin \varphi N_y)
           -d N_z. \label{nalpha}
\end{equation}
 $N_{x, y, z}$ are
entirely expressed in terms of spheroidal components as shown above.
In an analogous way we can obtain the remaining nonlinear parts 
$N_{\eta, \varphi}$.  

Using Eqs. (\ref{unit}) we transform the inverse of the penetration 
depth  tensor 
 given in Cartesian coordinates by a diagonal tensor with components
$(\lambda^{-2}_{ab}, \lambda^{-2}_{ab}, \lambda^{-2}_c)$, 
(recall Eq. (\ref{linear})) to spheroidal coordinates:
\begin{mathletters} 
 \[ \tilde{\Lambda}^{-1} = \lambda^{-2}_{ab}
\left[ \begin{array}{ccc} \rho_1 & \rho_2 & 0 \\
                           \rho_2 & \rho_3 & 0 \\
                           0 & 0 & 1\end{array} \right] \]
\end{mathletters}
where $\rho_{1, 2, 3}$ are defined by
\begin{mathletters}
\label{tensor}
\begin{equation}
\rho_1=\frac{\xi ^2 (1-\eta^2) +\delta^{-1} (1+\xi^2) \eta ^2}
        {\xi^2+\eta^2},
\end{equation}
\begin{equation}
\rho_2=-\frac{(1-\delta^{-1}) \xi (1+\xi^2)^{1/2} \eta (1-\eta^2) 
^{1/2} }
        {\xi^2+\eta^2},
\end{equation}
\begin{equation}
\rho_3=\frac{\delta^{-1} \xi ^2 (1-\eta^2) + (1+\xi^2) \eta ^2}
        {\xi^2+\eta^2}.
\end{equation}
\end{mathletters}
The resulting form of the dimensionless equations in 
spheroidal coordinates (which we will solve numerically) is
\begin{mathletters}
\label{eqsp}
\begin{equation}
f^2( \nabla\times\nabla\times {\bf V})_{\xi} =
-f^2((\rho_1 V_{\xi}+\rho_2 V_{\eta})-N_{\xi}), \label{eqspa}
\end{equation}

\begin{equation}
f^2( \nabla\times\nabla\times {\bf V})_{\eta} =
-f^2((\rho_2 V_{\xi}+\rho_3 V_{\eta})-N_{\eta}), \label{eqspb}
\end{equation}

\begin{equation}
f^2( \nabla\times\nabla\times {\bf V})_{\varphi} =
-f^2( V_{\varphi}-N_{\varphi}). \label{eqspc}
\end{equation}
\end{mathletters}
Expressions for  the differential operator
$f^2 \nabla\times\nabla \times {\bf V}$ in spheroidal coordinates are
included in Appendix \ref{other}. Equations (\ref{eqsp}) have to supplemented
with appropriate boundary conditions, as discussed in Section
II. The boundary condition at infinity is satisfied by the use of
the analytic solution for ${\bf R}^3 \setminus \Omega$.
Since $\lambda_{ab} \ll C$ we can
put  ${\bf V}\equiv 0$ at $\xi=0$,
as all the fields vanish deep
inside the sample. Continuity of the ${\bf {\cal H}}$ field on $\partial 
\Omega$ is achieved through
\begin{equation}
\nabla\times {\bf V}={\bf {\cal H}} \mid _{\xi=\xi_0},
\label{lon2}
\end{equation}
where the right hand side is the external dimensionless field at the
surface of the ellipsoid.
The remaining boundary  condition  ${\bf J \cdot n}$ $\equiv
J_{\xi}=0$ on $\partial \Omega$ is generally nonlinear and readily
enforced by observing that the RHS of Eq. (\ref{eqspa})  is $\propto
J_{\xi}$.

\subsection{Computational Grid and Discrete Variables}
\label{cgrid}

The implementation of the perturbation
method from Section \ref{comp} is not restricted to a particular 
algorithm for
solving the relevant equations in $\Omega$. In the remaining part of
this section
we outline as an example one suitable algorithm using a modification of 
the Gauss-Seidel relaxation method.
We first discuss the discretization of the complete
nonlinear, three-dimensional (3D) problem. It is then possible to consider,
as a special case, the   two-variable discretization of the  linear
problem were the $\varphi$ dependence is known analytically. The 
numerical solution to such a problem is then used as
the initial guess for the relaxation method of the complete 3D problem.

Eqs. (\ref{eqsp}) or their counterparts (\ref{maxlon}) and (\ref{jnl}),
have definite parity: $v_x$, $v_y$ are even and $v_z$ is odd in $z$. It 
is therefore sufficient to consider only the upper, {\it positive z} 
($\Omega_{+}$), or the lower,  {\it negative z}
($\Omega_{-}$) half of $\Omega$ and extend by parity the obtained solution 
to the whole $\Omega$.
The computational domain $G$ is obtained by parameterizing the physical
domain $\Omega_-$ using  oblate spheroidal coordinates ($\xi$, $\eta$,
$\varphi$).
We consider a uniform grid on $G$ with
mesh widths $h_{\xi}$, $h_{\eta}$ and $h_{\varphi}$.
The choice of a grid uniformly spaced in variable $\eta$,
generates denser grid points corresponding to the part of
$\Omega_-$ (the $\eta \approx 0$ region) with higher curvature and
greater field variation. An arbitrary grid point on $G$ is given by
 ($\xi_i$, $\eta_j$, $\varphi_k$) or just $(i, j, k)$ for brevity:
\begin{equation}
{\bf x}_{i,j,k}=\xi_i \hat{\xi}+\eta_j \hat{\eta}+ \varphi_k
\hat{\varphi}, \label{point}
\end{equation}
where the grid coordinates are given by
\begin{equation}
\xi_i=i h_\xi, \: \eta_j=-1+(\frac{1}{2}+j) h_\eta, \:
\varphi_k=-\pi+k h_\varphi, \label{coord}
\end{equation}
and the indices run through values $i=0,.., n_{\xi}$,
$ j=0,.., n_{\eta}$ and $k=0,.., n_{\varphi}$. Mesh widths are given by
$h_\xi=\frac{\xi_0}{n_\xi}$, $h_\eta=\frac{2}{2 n_\eta+1}$ and
$h_\varphi=\frac{2 \pi}{n_\varphi}$.

The discrete variables are denoted by the same symbol
as their continuous counterparts, for example, ${\bf V}_{\xi; i, j, 
k}$
represents ${\bf V}_\xi$ at the grid point  ($\xi_i$, $\eta_j$,
$\varphi_k$). The discretized approximations of derivatives used have
second order accuracy in the mesh widths. We will use the letter $D$ to
represent the discretized approximation, upper indices $0, +, -$ denote
the central, forward and, backward approximation respectively, and 
the
 lower indices denote the  corresponding variables of differentiation.
For example, $D^{0+}_{\xi \eta}$ denotes the mixed differentiation with
respect to $\xi$ and $\eta$, where the central difference formula is used 
in $\xi$
and the forward difference formula in $\eta$ variable. For the interior
grid points we only use the central difference formula for all the
derivatives and omit the upper indices.
We also use central differences for derivatives at all the
grid points on $\partial G$ that do not require introduction of
fictitious grid points $\notin G$. For example, in computing $D^0_\xi$ 
at
$(n_\xi, j, k)$ we would need to use a grid point at $i=n_\xi+1$ 
$\notin$
$G$, we avoid that by using a backward difference $D^-_\xi$ at 
$(n_\xi, j,
k)$. Similarly we employ forward differences were necessary.

\subsection{Implementation of Boundary Conditions and Equations on $G$}
\label{bcon}
The grid boundary, $\partial G$, consists of six two-dimensional planar
surfaces with grid points described by
$(0, j, k)$, $(n_\xi, j, k)$, $(i, 0, k)$, $(i, n_\eta, k)$, $(i, j, 
0)$
and $(i, j, n_\varphi)$.

On the surface $\xi=0$, $(0, j, k)$, which
corresponds to the equatorial $(z=0)$ disk in $\Omega$ with radius 
equal
to the focal length $f$, we impose trivial Dirichlet boundary
conditions. As we have discussed in Section II, this follows from
 the requirement that deep inside $\Omega$
all fields should vanish. This eliminates  possible difficulties
from the singularities of the various differential operators at 
$\xi=0$.
Any remaining singularities of the equations on $G$ would come from    
 points  with coordinates
$\eta=0$ and $\eta=\pm 1$. On the surface  $\eta=0$ $(i, j=n_\eta, k)$,
corresponding to part of the $z=0$ plane, 
we have from the known parity of ${\bf V}$:
\begin{mathletters}
\label{parity}
\begin{equation}
 V_{\xi, \varphi}=0 |_{\eta=0}, \label{paritya}
\end{equation}
\begin{equation}
\partial_\eta V_{\eta}=0 |_{\eta=0}. \label{parityb}
\end{equation}
\end{mathletters}
We implement Eq. (\ref{parityb}) as $D^-_\eta V_{\eta; i, j=n_\eta, 
k}=0$:
\begin{equation}
D^-_\eta V_{\eta; i, j, k}=\frac{1}{2 h_\eta} (3 V_{\eta; i, j, k}
-4 V_{\eta; i, j-1, k}+ V_{\eta; i, j-2, k}), \label{back}
\end{equation}
and in the iterative solution we write down explicitly $V_{\eta; 
i,j,k}$ from this expression.
The remaining region on  $G$ that could result in 
singularities  of the differential operators is at
$\eta=-1$, which corresponds to a line through the ``south pole" and the
origin of $\Omega$,
i.e., a segment starting at  the origin and ending at a point
with  Cartesian coordinates $(0, 0, z)$. The choice of  grid 
given in
Eq. (\ref{coord}) excludes this segment, the closest point on the grid 
is $h_\eta$/2 away. The numerical solution for ${\bf V}$ in the vicinity of
$\eta=-1$ is well behaved (as it is for the linearized
equations in the geometries that permit an analytic solution in
$\Omega$). It is therefore possible to extrapolate
 the obtained numerical solution to $\eta=-1$. At the grid boundary
surface  $(i, j=0, k)$ we use the forward difference formula for the
derivatives with respect to $\eta$, $D^+_\eta$ is obtained analogous to
Eq. (\ref{back}) by replacing $h_\eta$ with $-h_\eta$ and the indices 
$j-1$,
$j-2$ by $j+1$ and $j+2$ respectively.
The second derivative $D^{++}_{\eta \eta}$ is taken as:
\begin{equation}
D^{++}_{\eta \eta}F_{i,j,k}=\frac{1}{h_\eta ^2}(-F_{i,j+3,k}+4
F_{i,j+2,k}-5 F_{i,j+1,k}+2 F_{i,j,k}), \label{2forw}
\end{equation}
where $F$ represents any component of a vector field. We can  obtain
similar formulae for $D^{0+}_{\xi \eta}$ and $D^{+0}_{\eta
\varphi}$.
At the  surface boundary $(i, j, k=0)$, where $\varphi=-\pi$, we 
proceed in an
analogous way, the derivatives with respect to $\varphi$ are expressed 
with forward differences. The other part of $\partial G$  with
$\varphi=const$ i.e., $(i, j, k=n_\varphi)$ corresponds to the same
boundary surface ($\varphi=-\pi$) and we can impose the simple periodic
boundary conditions
\begin{equation} 
F_{i,j,n_\varphi}=F_{i,j,0}. \label{period}
\end{equation}
On the remaining part of $\partial G$, the surface $(i=n_\xi, j, k)$ at
$\xi=\xi_0$, we impose continuity (as discussed in section IV) of
the $\eta$ and $\varphi$ components of
${\bf {\cal H}}$. From Eq. (\ref{lon2})
we can express
$V_{\eta;i,j,k}$  and $V_{\varphi;i,j,k}$ respectively to obtain
their updated values in each step of the relaxation procedure. The
equation for $V_{\xi; i,j,k}$ is obtained from ${\bf J} \cdot {\bf
n}$=0 by setting the RHS of Eq. (\ref{eqspa}) to zero
\begin{equation}
\rho_1 V_\xi + \rho_2 V_\eta - N_\xi =0. \label{zero}
\end{equation}
In the relaxation procedure, the nonlinear term $N_\xi$ is included
using Picard's method [\onlinecite{pic}], the value of $N_\xi$ is taken from the
previous iteration. If we denote by an upper index $n$  the number of
the iteration (in the relaxation procedure), the nonlinear boundary     
 condition can be implemented as
\begin{equation}
V^{n+1}_\xi=-\frac{\rho_2}{\rho_1} V^{n+1}_\eta+\frac{1}{\rho_1}
N^{n}_\xi.  \label{picard}
\end{equation}
The  nonlinear and nonanalytic
terms can be simply
included by using Picard's method.
 In addition to the 
nonlinear boundary condition (\ref{picard}) we shall 
 also use  Picard's method to include the nonlinearities
stemming from Eq. (\ref{eqsp}).

For the grid points $\notin$ $\partial G$ it is possible
to use  central differences. In the LHS of Eq. (\ref{eqsp}), given
explicitly in  Appendix C, we replace each partial derivative
 by the appropriate central difference, $D$.

\subsection{Modified Gauss-Seidel Relaxation}
\label{seidel}
The solution to the nonlinear, 3D problem using
the relaxation method consists of two steps. The first is obtaining
a good initial guess using the numerical solution to the linear 
equations
given by Eq. (\ref{maxlon}) and (\ref{linear}). The $\varphi$
dependence is then known and it can be separated out.
The method of successive overrelaxation can be applied to the
resulting two-variable problem in the coordinates $\xi$ and $\eta$.
The boundary conditions on the continuity of ${\bf{\cal H}}$, as discussed in 
the previous subsection,  are implemented:
 the magnetic field outside is taken in the $\epsilon=0$ limit
($A_{1\|}=A_{1\|}(0)$).
In the relaxation procedure each component of ${\bf V}$ is  expressed in
terms of  the corresponding component of the linearized
($N_{\xi,\eta,\varphi}=0$), two-dimensional form of Eq.  (\ref{eqsp})
so that the resulting matrix equation is
diagonally dominant [\onlinecite{pde}].  For a detailed discussion of the
method see  [\onlinecite{rec,smith}]. The numerical solution, the
distribution of $V(\xi_i, \eta_j)$, is
supplemented with the known $\varphi$ dependence and $m_\|$
is calculated using the perturbation method.

The second part of the algorithm is a modification of the Gauss-Seidel 
(the overrelaxation parameter, $\omega=1$) method
for the full 3D problem. The previously obtained 
solution for the
linearized equations is the initial guess for $V_{i,j,k}$,
and the value $m_\|$ i.e., the corresponding $A_{1\|}$,  is used in the
expression for the ${\bf {\cal H}}$ outside. If we denote the
linear part of Eq. (\ref{eqsp}) as ${\bf L(V)}$ and the nonlinear
part as ${\bf N(V)}$ the relaxation procedure can be described
symbolically as:
\begin{equation}
{\bf L(V}^{(n+1)}{\bf)}={\bf N(V}^{(n)}{\bf)}, \qquad \qquad n=1,2,..
\label{iter}
\end{equation}
After completion of each relaxation step we update the old values, ${\bf
V}^{(n)}$, at each grid point as ${\bf N(V}^{(n)})={\bf N(V}^{(n+1)})$.
As done with the boundary condition from Eq. (\ref{picard}) we use 
Picard's method to include the nonlinear terms which are also
nondifferentiable.
The numerical solution for the linearized equations
is a very good initial guess for the full problem, since the nonlinear 
terms are small, which compensates for the relatively slow convergence of     
Picard's method.
                              
\section{Numerical Tests}
\label{tests}
For our computations we used the Cray C90 of the  Minnesota 
Supercomputer Institute.
We first tested the computer code on the example of the
linear, isotropic sphere with applied field along the $x$ axis, 
where the analytic solution is known.
The spherical geometry was realized as the spherical limit of the
spheroid. We  used $\xi_0=1000$ and $f=0.1$, so that
the radius of  sphere was $a \rightarrow f \xi_0=100$ (in units of 
$\lambda$),
the corresponding  $\epsilon=\lambda/ a\: = \:10^{-2}$ and the ratio
of the spheroidal semi-axes is $A/C=100.005/100$. We
used the two-variable version of the code in the variables $\xi$ and 
$\eta$. The computational grid  spanned a
spherical shell of thickness 7 $(\lambda)$,  and because  fields
 decay exponentially away from the surface, we imposed
 trivial Dirichlet conditions at $\xi=930$.  We  used $n_\xi=200$
and $n_\eta=50$ for  the number of grid points. To test the 
convergence of the relaxation algorithm, we tried the very poor         
 initial guess of zero fields everywhere on $G$. The boundary           
conditions were taken in the $\epsilon=0$ limit, that is 
$A_{1\|}=A_{1\|}(\epsilon=0)$
from Eq. (\ref{a1}). The overrelaxation parameter was $\omega=1.8$
and after 1000 relaxation steps (43 sec of CPU time) we obtained
 ${\bf m}$ using Eq. (\ref{magmoms}) and the perturbation method
with Eq. (\ref{ms}).
In the term ${\bf m_2}$ the distribution of ${V_\eta, \: V_\varphi}$ at
$\xi=\xi_0$ (from the numerical solution on $G$) 
was extended by parity to the entire $\partial \Omega$
and supplemented with the known $\varphi$ dependence. Thus,
integration over $\varphi$ was performed analytically and that with respect
to $\eta$, numerically. The extracted constant (from Eq. (\ref{ms})) was
$\alpha_\|=3.1$, within $3 \: \%$ of the
analytically obtained value of $3$ from Eq. (\ref{exact}).
The use of a conventional procedure
to compute $m$ would require repeating the whole iteration procedure  to
get an improved value for $\alpha_\|$, as we have in principle 
described in \ref{per}. 

We also verified that the numerical solution for  
the full three-variable  problem in this geometry and exact boundary    
conditions, has the correct form. The angular dependence of the
solution was $J_\eta$, $V_\eta$ $\propto \eta \sin \varphi$
and  $J_\varphi$, $V_\varphi$ $\propto (1-\eta ^2)^{1/2} \cos \varphi$
which in the spherical limit $\eta \rightarrow \cos \theta $
 corresponds to the analytical solution for a
sphere. The numerical solution, for the range of grids considered, was
accurate between three and four significant figures for every grid
point where ${\bf J}$ $({\bf V})$ was numerically significant.

For the physical results related to the nonlinear response of an
oblate spheroid we refer to [\onlinecite{zv}], and discuss here
only some aspects not covered there, that illustrate the numerics.
 Analysis of the transverse magnetic moment shows that 
[\onlinecite{zv}]
\begin{equation}
m_\bot \propto \frac{H_a}{H_0} H_a F(\Psi), \label{quadratic}
\end{equation}
from which we infer that $\alpha_\bot \propto H_a / H_0 \: F(\Psi)$,
where $F(\Psi)$ is the  angular dependence on $\Psi$ which
has [\onlinecite{zv}] period $\pi/2$. We
will give results for $\Psi=\pi /8$, approximately the maximum of $F$.
The longitudinal moment, due to  field penetration, differs from $m_0$ 
and it can be characterized, as we have shown earlier, by the parameter
 $\alpha_\|$. This parameter 
includes contributions from the linear part of ${\bf j(v)}$,
independent of $H_a$, and from the nonlinear part,
dependent both on $H_a/H_0$ and on $\Psi$. 

We consider an oblate 
spheroid with $\xi_0=0.144338$ $(A/C=7)$ at $H_a/H_0=0.1$ (in the 
experimentally relevant range) and
$\Psi=\pi /8$. For $f=1000$ (in units of $\lambda_{ab}$, defined in \ref{jvv} 
 and \ref{dform})
 we have used $n_\xi=550$, $n_\eta=50$ and  $n_\varphi=30$.  The
results for $\alpha_\|$ and $\alpha_\bot$ are given as a function of 
the material parameter $\delta=(\lambda_c / \lambda_{ab})^2$ in Table I.
For fixed $\xi_0$ (fixed shape), we have
considered various sizes of spheroid (different $f$), changing $\epsilon
\equiv \lambda_{ab}/C \ll 1$ by a
factor of four (at fixed $\lambda_{ab}, \lambda_c$) and verified that
$\alpha_\|$, $\alpha_\bot$ are size independent within  numerical
accuracy.

In the next two figures we display some of the numerical results for 
the field distributions calculated under the same conditions and with 
the same parameter values as in the previous paragraph. In Fig. 3, we   
show results for the  current at surface of the spheroid,  ($\xi=\xi_0$).
The components $J_{\xi, \eta}(\xi_0, \eta, \varphi)$, at the fixed azimuthal  
angle $\varphi=48^\circ$, are shown  as functions of $\eta$. For comparison  
 we recall here 
also the corresponding angular behavior for a sphere, as given earlier 
in this Section. In Fig. 4 we show some results for the magnetic field 
as it penetrates into the sample. We plot the two  components 
${\cal H}_{\eta, \varphi}(\xi, \eta, \varphi)$ as functions of $\xi_0-\xi$ at 
constant $\varphi=0^\circ$, $\eta=-0.693$, (see Fig. 2).
The plot illustrates the difference between the components of the field 
arising purely from the linear equations and those which are due to the 
nonlinear effects.
The component ${\cal H}_\eta$ (at $\varphi=0^\circ$) is very predominantly 
``linear'', 
and displays exponential-like behavior. The other component plotted, 
${\cal H}_\varphi$, vanishes in the linear case (at $\varphi=0^\circ$), and it 
arises solely from the nonlinear effects. This behavior is far from being an 
exponential; its derivative changes sign, for the same physical reasons as the 
nonlinear current does, as discussed in [\onlinecite{zv}].
                                                               
\section{Conclusions and Future Work}
\label{future}
In this paper, as the main result, we have presented a perturbation
method to compute the magnetic moment of a bulk nonlinear and 
anisotropic superconductor.
This method could be implemented
in conjunction with various algorithms for  solving
boundary value problems in electrodynamics. Suitable generalizations
would certainly increase the range of its applicability from that discussed in 
this paper.  Obvious examples include
 considering in detail other shapes of $\Omega$ and computing higher
order multipoles. We have showed that our method increases the
accuracy of computation while very significantly reducing the required
computational work.

In this paper the numerical example of
an oblate spheroidal geometry was discussed in detail to illustrate the
perturbation method and also to give guidelines for  possible
improvements. The numerical algorithm which was employed for solving
the nonlinear Maxwell-London equations provided more than sufficient    
 accuracy, as seen from experimental considerations: the uncertainty of 
 the input experimental parameters significantly exceeds the accuracy  
of the results obtained. For our computations we have used Cray C90 and   
memory requirements were not the limiting factor. It is however possible to 
 make  various improvements to the numerical algorithm.  
 One could consider nonuniformly spaced grid points
 along the $\xi$ coordinate,
denser close to the $\xi=\xi_0$, in order to  reduce the overall number 
of grid points. It might also be advantageous to replace the Gauss-Seidel 
relaxation on
$G$ by some other method, such as GMRES [\onlinecite{saad}] or one of the
various multigrid algorithms [\onlinecite{rec}].  In
future work we will  consider some of these improvements and
investigate possible generalizations of the perturbation method
presented in this paper.

\acknowledgments
We thank A. Bhattacharya, A.M. Goldman, B. Bayman and B.P. Stojkovi\'c
 for discussions concerning 
experimental and theoretical aspects of our work and B. F. Schaudt 
 for help with preparing the figures. I.
\v{Z}. acknowledges support from the Stanwood Memorial Fellowship.

\appendix
\section {The Nonlinear Relation ${\bf \lowercase{j(v_s)}}$}
\label{jofv}
In order to express ${\bf j(v_s)}$ from Eq.
(\ref{nonlinjv}), we  introduce two
coordinate systems: $x-y$, fixed in space such  that the
applied field is along the $x$-axis, and $x'-y'$ which is fixed to the 
crystal.
We consider an order parameter of the so called $d$-wave form:
\begin{equation}
\Delta=\Delta_0 \sin(2 \phi),
\label{op}
\end{equation}
where $\phi$ is the azimuthal angle referred to a node and $\Delta_0$
the gap amplitude.
It has been shown [\onlinecite{zv}] that in the field range of experimental 
interest and with suitable assumptions for the Fermi surface, Eq. 
(\ref{nonlinjv}) can be rewritten, at sufficiently low temperatures as
\begin{mathletters}
\label{jnl}
\begin{equation}
j_{x',y'}  =-e\rho_{ab} v_{x',y'} (1-\frac{t_1}{ v_c}\left|v_{x',y'}
\right|),
\end{equation}
\begin{equation}
j_z =-e\rho_{c} v_{z} (1-\frac{t_2}{ v_c}\frac{v_{x'}^2+v_{y'}^2}
{\left|v_{x'}\right|+\left|v_{y'}\right|}),
\end{equation}
\end{mathletters}
where $t_1$ and $t_2$ are constants, $t_1=\frac{9\pi}{64}$,
$t_2=\frac{3\pi}{32}$.  The  critical velocity is $v_c=\Delta_0/v_f$.
We define here $\rho_{ab}=\frac{c^2}{4 \pi e^2} \lambda^{-2}_{ab}$ and
$\rho_{c}=\rho_{ab}\frac{\lambda^2_{ab}}{\lambda^2_c}$.

\section{Quantities in spheroidal coordinates}
\label{sphero}
\subsection{Magnetic Moment}
\label{mom}
 We show here that the expression for the longitudinal magnetic moment
for an oblate spheroid, given by Eq. (\ref{longmag}), is equivalent to 
the standard result expressed in terms of the demagnetization
factors. For a field applied along the $x$ axis, 
in the limit $\tilde{\Lambda}=0$, we have $\:$ [\onlinecite{landau}]:
\begin{equation}
m_{0\|}\equiv m_{0x}=-\frac{H_a V}{4 \pi (1-n_x)}, \label{mx}
\end{equation}
where V is the volume of a spheroid and
\begin{equation}
n_x=-\frac{1}{2}-\frac{1+e^2}{2 e^3}(e-arctan(e)) \label{nx}
\end{equation}
is the appropriate demagnetization factor. The eccentricity
is  given by
$e=[A^2/C^2-1]^{1/2}=1/ \xi_0$. In terms of spheroidal coordinates we
have $A=f (1+\xi_0^2)^{1/2}$, $C=f \xi_0$ and $V=\frac{4 \pi}{3} f^3
(1+\xi_0^2) \xi_0$. Including these expressions in Eq. (\ref{mx}) we
get
\begin{equation}
m_{\|}=-\frac{2}{3} f^3 H_a \frac{\xi_0}{1+1/(1+\xi_0^2)-
\xi_0 arctan(1/\xi_0)} \label{mapp}
\end{equation}
 and we recover $m_{\|}=\frac{2}{3} f^3 A_1$.

\subsection{Integral and Differential Operators}
\label{other} 

The metric coefficients in oblate spheroidal coordinates are given by 
[\onlinecite{magnus}]:
\begin{mathletters}
\label{metric}
\begin{equation}
g_{11}=f^2 \frac{\xi^2+\eta^2}{1+\xi ^2},
\end{equation}
\begin{equation}
g_{22}=f^2 \frac{\xi^2+\eta^2}{1-\eta ^2},
\end{equation}
\begin{equation}
g_{33}=f^2 (1+\xi ^2) (1-\eta ^2).
\end{equation}
\end{mathletters}
One can  then calculate the appropriate
elements of integration and differential operators. For example,
\begin{equation}
\int_{\partial \Omega} dS =\int_{-1} ^{1} d\eta \int_0 ^{2 \pi} 
d\varphi f^2
(1+\xi^2)^{1/2} (\xi ^2+ \eta ^2)^{1/2}, \label{ds}
\end{equation}
\begin{equation}
\int_{ \Omega} d\Omega =\int_0^{\xi_0} d\xi\int_{-1} ^{1}
d\eta \int_0 ^{2 \pi} d\varphi f^3 (\xi ^2+ \eta ^2). \label{domega}
\end{equation}
$\xi_0$ corresponds to the value of $\xi$ at the boundary $\partial
\Omega$. The gradient is
\begin{equation}
 {\bf \nabla}=\frac{(1+\xi^2)^{1/2}}{f(\xi^2+\eta^2)^{1/2}} \hat{\xi}
\partial_\xi
+\frac{(1-\eta^2)^{1/2}}{f(\xi^2+\eta^2)^{1/2}} \hat{\eta} 
\partial_\eta
+\frac{1}{f (1+\xi^2)^{1/2} (1-\eta^2)^{1/2}} \hat{\varphi}
 \partial_\varphi. \label{nabla}
\end{equation}

To solve equations (\ref{maxlon}) the expression
$\nabla\times\nabla\times {\bf v}$ should be
transformed to oblate spheroidal coordinates:
\begin{eqnarray}
f^2 (\nabla\times\nabla\times {\bf v})_{\xi} &=
a_0\partial_{\eta\eta}v_{\xi}+
a_1\partial_{\eta}v_{\xi}+
a_2\partial_{\varphi\varphi}v_{\xi}+
a_3v_{\xi}+
a_4\partial_{\xi\eta}v_{\eta}+
a_5\partial_{\xi}v_{\eta}+
a_6\partial_{\eta}v_{\eta} \nonumber \\
&+a_7v_{\eta}
+a_8\partial_{\xi\varphi}v_{\varphi}+a_9\partial_{\varphi}v_{\varphi},
\end{eqnarray}

\begin{eqnarray}
f^2 (\nabla\times\nabla\times {\bf v})_{\eta} &=
b_0\partial_{\xi\eta}v_{\xi}+
b_1\partial_{\xi}v_{\xi}+
b_2\partial_{\eta}v_{\xi}+b_3v_{\xi}+
b_4\partial_{\xi\xi}v_{\eta}+
b_5\partial_{\xi}v_{\eta}+
b_6\partial_{\varphi\varphi}v_{\eta} \nonumber \\
&+b_7v_{\eta}+
b_8\partial_{\eta\varphi}v_{\varphi}+b_9\partial_{\varphi}v_{\varphi},
\end{eqnarray}

\begin{eqnarray}
f^2 (\nabla\times\nabla\times {\bf v})_{\varphi} &=
p_0\partial_{\xi\varphi}v_{\xi}+
p_1\partial_{\varphi}v_{\xi}+
p_2\partial_{\eta\varphi}v_{\eta}+
p_3\partial_{\varphi}v_{\eta}+
p_4\partial_{\xi\xi}v_{\varphi}+
p_5\partial_{\eta\eta}v_{\varphi} \nonumber \\
&+p_6\partial_{\xi}v_{\varphi}+
p_7\partial_{\eta}v_{\varphi}+p_8 v_{\varphi}.
\end{eqnarray}
We recall that $f$ is the  focal length scale factor.
The coefficients $a_i,b_i,p_i$ are given by (using the abbreviations
 $u\equiv (1+\xi^2)^{1/2}$,  $s\equiv (1-\eta^2)^{1/2}$,
$w\equiv(\xi^2+\eta^2)^{1/2}$):
\begin{eqnarray}
&a_0=-\frac{s}{u}a_4=\frac{s}{u}b_0=-\frac{s^2}{u^2}b_4
=\frac{s^2}{u^2}p_4=p_5=-\frac {s^2}{w^2}, \nonumber \\
&a_1=-\frac{\eta}{\xi}b_5=\frac{\eta}{\xi}p_6
=-p_7=-\frac{2 \eta}{w^2}, \nonumber \\
&a_2=-\frac{w}{\xi s}a_9=b_6
=-\frac{w}{u \eta}-p_8=-\frac{1}{u^2 s^2}, \nonumber \\
&a_3=\frac{2 \eta^2-\xi^2+3 \xi^2 \eta^2}{w^6}, \nonumber \\
&a_5=-\frac{u^2 \eta}{\xi s^2}a_6=\frac{u^2}{s^2}b_1
=-\frac{u^4 \eta}{\xi s^4}b_2=\frac{u^3 \eta }{s w^4}, \nonumber \\
&a_7=-\frac{\xi u \eta(3+\xi^2-2\eta^2)}{s w^6 }, \nonumber \\
&a_8=-\frac{u}{s}b_8=p_0=\frac{u}{s}p_2=\frac{1}{s w}, \nonumber \\
&b_3=\frac{\xi \eta s(3+2 \xi^2-\eta^2)}{u w^6}, \nonumber \\
&b_7=-\frac{2\xi^2-\eta^2-3 \xi^2 \eta^2}{w^6}, \nonumber \\
&p_1=-\frac{\xi s^3}{u \eta}p_3=-\frac{\xi s}{u^2  w^3}. \nonumber \\
\end{eqnarray}

\subsection{Magnetic field which contributes to ${\bf m}$}
\label{field}
We write down  here explicitly the contributions to the magnetic        
 field which arise from the potential $\Phi$ given by the sum of 
  (\ref{phiout}) and (\ref{phiout2}). From Eq. (\ref{outa}) and 
(\ref{nabla}) we have
\begin{mathletters}
\label{he}
\begin{equation}
H_\xi=\frac{(1-\eta ^2)^{1/2} } {(\xi^2+\eta^2)^{1/2}}
(f_1(\xi) \cos \varphi+f_1\bot(\xi) \sin \varphi),
\label{hea}
\end{equation}
\begin{equation}
H_\eta=-\frac{\eta}{(\xi^2+\eta^2)^{1/2}}
(f_2(\xi) \cos \varphi+f_2\bot(\xi) \sin \varphi),
\label{heb}
\end{equation}
\begin{equation}
H_\varphi=-\frac{1}{(1+\xi^2)^{1/2}}
(f_2(\xi) \sin \varphi-f_2\bot(\xi) \cos \varphi),
\label{hec}
\end{equation}
\end{mathletters}
where the functions $f_1(\xi)$, $f_2(\xi)$ are given by:
\begin{mathletters}
\label{ff}
\begin{equation}
f_1(\xi)=H_a \xi +A_1(1+\frac{1}{1+\xi ^2}-\xi arctan(1/\xi)),
\label{ffa}
\end{equation}
\begin{equation}
f_{1\bot}(\xi)=A_{1\bot}(1+\frac{1}{1+\xi ^2}-\xi arctan(1/\xi)),
\label{ffb}
\end{equation}
\begin{equation}
 f_2(\xi)=H_a (1+\xi^2)^{1/2} +A_1(\frac{\xi}{(1+\xi ^2)^{1/2}}-
(1+\xi^2)^{1/2} arctan(1/\xi)),
\label{ffc}
\end{equation}
\begin{equation}
 f_{2\bot}(\xi)=A_1{\bot}(\frac{\xi}{(1+\xi ^2)^{1/2}}-
(1+\xi^2)^{1/2} arctan(1/\xi)).
\label{ffd}
\end{equation}
\end{mathletters}
Terms with $f_1$ and $f_2$ represent the longitudinal part
of ${\bf H}$ and those
with $f_{1\bot}$ and $f_{2\bot}$ the transverse part.

\section{Evaluation Of ${\bf {\bar {\lowercase{m}}}_2}$ to ${\bf O(\epsilon 
\lowercase{m}_0)}$  
for a Sphere}
\label{eva}
We evaluate here the magnitude $\bar {m}_2$ for the example of an 
isotropic spherical superconductor with a linear ${\bf j(v_s)}$ 
relation. We take the field along the $z$ direction, since the magnetic 
moment $m_2$ is independent of this choice. 
As explained in the text, we must find the internal fields by solving 
(\ref{maxlon}) (which in this case reduces to the Helmholtz equation) 
with boundary conditions enforcing continuity of $H_\varphi$ and 
$H_\theta$ and external fields calculated in the $\epsilon=0$ limit.     
By symmetry, $H_\varphi \equiv 0$ and we need only to impose the continuity 
  of $H_{\theta}$ at the boundary, $r=a$. We obtain:  
\begin{equation}
H_{\theta} \equiv -A\frac{\lambda ^2}{r ^3} 
((1+\frac{r^2}{\lambda^2}) 
\sinh(r/\lambda)-\frac{r}{\lambda}\cosh(r/\lambda))\sin 
\theta=(-H_a+\frac{m_0}{r^3}) \sin \theta.
\label{heta}
\end{equation}
A is a constant to be determined and $m_0$ is given by Eq. (\ref{mos}). 
Since $\epsilon=\frac{\lambda}{a} \ll 1$ we can approximate 
$\sinh(a/\lambda) \approx \cosh(a/\lambda) \approx 
\frac{1}{2}e^{(a/\lambda)}$.
Keeping only the leading term in the LHS of Eq. (\ref{heta}) we get
\begin{equation}
A=3 H_a e^{-(a/\lambda)}
\end{equation}
and
\begin{equation}
H_\theta=-\frac{3}{2} H_a \frac{a}{r} e^{-(a-r)/\lambda} \sin \theta.
\end{equation}
We can now compute
\begin{equation}
{\bar {\bf m}}_2=\frac{1}{8 \pi} \int_{\Omega} d \Omega {\bf H}
\end{equation}
Integration is performed in spherical coordinates giving
\begin{equation}
m_2=\frac{1}{2} \lambda a^2 H_a=-\frac{\lambda}{a} m_0 \sim O(\epsilon 
m_0).
\end{equation}
This result (O($\frac{\lambda}{a} m_0$)) was expected since for 
exponentially decaying fields, 
integration over the whole volume of $\Omega$ is effectively only
integration over the region $\sim \lambda$ away from its surface.
 

%
%
\begin{figure}
\caption{Geometry 
considered here. The superconducting region $\Omega$ is an oblate
ellipsoid of revolution.  The $x$, $y$ and $z$ directions
are fixed in space. The field is applied along the $x$ axis, as 
indicated, while ${\bf m_\bot}$ is along the $y$ axis. The long and
short semiaxes values are called 
$A$ and $C$ in the text, respectively.}
\label{fig1}
\end{figure}
\begin{figure}
\caption{The relation between oblate spheroidal $(\xi, \eta, \varphi)$ 
and  Cartesian coordinates is illustrated at a fixed azimuthal angle 
$\varphi=0^\circ$. The surfaces $\xi=constant$ are oblate ellipsoids of 
revolution around the $z$-axis with semiaxes $A=f (1+\xi^2)^{1/2}$, 
 $C=f \xi$. The surfaces $\left| \eta \right|=constant$ are one-sheeted 
hyperboloids of revolution. The surfaces $\varphi=constant$ are planes 
through the $z$-axis making an angle $\varphi$ with the $x-z$ plane.}  
\label{fig2}
\end{figure}  
\begin{figure}
\caption{The spheroidal components of the dimensionless current (see 
(\protect{\ref{dimless}})) at the  surface of the  spheroid $(\xi=\xi_0)$. The 
quantities plotted are $J_\eta (\xi_0, \eta, \varphi=48^\circ)$ (solid 
line) and $J_\varphi (\xi_0, \eta, \varphi=48^\circ)$ (dashed line). We 
have used $H/H_0=0.1$, $A/C=7$, $\Psi=\pi/8$, $\delta=16$.}
\label{fig3}
\end{figure}
\begin{figure}
\caption{Illustration of the nonlinear effects on the magnetic field. 
The solid line is the component ${\cal H}_\eta (\xi, \eta=-0.693, 
\varphi=0^\circ)$, which is very predominantly ``linear'', normalized to  
its maximum value, while the dashed line is ${\cal H}_\varphi (\xi, 
\eta=-0.693, \varphi=0^\circ)$, which arises solely from  nonlinear 
effects, also normalized to its own, much smaller, maximum value. Both 
components are plotted as functions of $D\equiv\xi_0-\xi$. The 
non-exponential behavior of the dashed line is a signature of its 
nonlinear character. All parameters used are as in the previous 
Figure.}
\label{fig4}
\end{figure}
%
%
%
\begin{table}
\caption{The parameters $\alpha_\|$, $\alpha_\bot$ which determine 
the magnetic moment for $\epsilon \neq 0$ (see Eq. (\protect{\ref{mmm}}), 
given as  functions of the material parameter
$\delta=(\lambda_c/ \lambda_{ab})^2$, computed for an oblate 
spheroid with $A/C=7$ at a field $H/H_0=0.1$.}
 \label{table1}
 \begin{tabular}{ccc}
$\delta=(\lambda_c/\lambda_{ab})^2$&$\alpha_\|$&$10^3$$\alpha_\bot$\\
\tableline
16&1.9&3.8\\
25&2.1&3.6\\
36&2.2&3.3\\
50&2.3&3.1\\
\end{tabular}
\end{table}

\end{document}